\documentclass[aps,prapplied,twocolumn,superscriptaddress,showkeys]{revtex4-2}

\usepackage{graphicx}
\usepackage{siunitx}
\usepackage{empheq}
\usepackage{amsmath}
\usepackage{amssymb}
\usepackage{txfonts}
\usepackage[colorlinks=true, linkcolor=blue,citecolor=red]{hyperref}
\usepackage{xcolor}
\usepackage{ulem,xpatch}

\DeclareSIUnit \belm {bm}
\DeclareSIUnit{\dBm}{\deci\belm}

\newcommand{\ii}{{\rm i}}
\newcommand{\ee}{{\rm e}}
\newcommand{\new}[1]{\textcolor{black}{#1}}

\newcommand{\be}{\begin{equaion}}
\newcommand{\bea}{\begin{eqnarray}}
\newcommand{\eea}{\end{eqnarray}}

\definecolor{green}{rgb}{0.1,.7,0.05}

\renewcommand\Re{\operatorname{Re}}

\newcommand{\calR}{{\mathcal R} }

\newcommand{\calS}{{\mathcal S}}
\newcommand{\calF}{{\mathcal F}}
\newcommand{\calA}{{\mathcal A}}

\newcommand{\calK}{{\mathcal K}}

\newcommand{\calL}{{\mathcal W}}

\begin{document}

\title{
Absolute determination of the single--photon optomechanical coupling rate via a Hopf bifurcation}
%Determination of the single--photon optomechanical coupling rate in time domain}
%\\
%Determination of the radiation pressure interaction rate in an optomechanical system}

\author{Paolo Piergentili}
	\affiliation{School of Science and Technology, Physics Division, University of Camerino, I--62032 Camerino (MC), Italy}
	\affiliation{INFN, Sezione di Perugia, I--06123 Perugia (PG), Italy}
\author{Wenlin Li}
	\affiliation{School of Science and Technology, Physics Division, University of Camerino, I--62032 Camerino (MC), Italy}
\author{Riccardo Natali}
	\affiliation{School of Science and Technology, Physics Division, University of Camerino, I--62032 Camerino (MC), Italy}
	\affiliation{INFN, Sezione di Perugia, I--06123 Perugia (PG), Italy}
\author{David Vitali}
	\email{david.vitali@unicam.it}
	\affiliation{School of Science and Technology, Physics Division, University of Camerino, I--62032 Camerino (MC), Italy}
	\affiliation{INFN, Sezione di Perugia, I--06123 Perugia (PG), Italy}
	\affiliation{CNR-INO, L.go Enrico Fermi 6, I-50125 Firenze, Italy}
\author{Giovanni Di Giuseppe}
	\email{gianni.digiuseppe@unicam.it}
	\affiliation{School of Science and Technology, Physics Division, University of Camerino, I--62032 Camerino (MC), Italy}
	\affiliation{INFN, Sezione di Perugia, I--06123 Perugia (PG), Italy}
\date{\today}

\begin{abstract}
We establish a method for the determination of the single--photon optomechanical coupling rate, which characterizes the
radiation pressure interaction in an optomechanical system. The estimation of the rate with which a mechanical oscillator, initially in a thermal state, undergoes a Hopf bifurcation, and reaches a limit cycle, allows us to determine the single–photon optomechanical coupling rate in a simple and consistent way. Most importantly, and in contrast to other methods, our method does not rely on knowledge of the system’s bath temperature and on a calibration of the signal. We provide the theoretical framework, and experimentally validate this method, providing a procedure for the full characterization of an optomechanical system, which could be extended outside cavity optomechanics, whenever a resonator is driven into a limit cycle by the appropriate interaction with another degree of freedom.
\end{abstract}

%\begin{abstract}
%	We establish a new method for the determination of the single--photon optomechanical coupling rate, which characterizes the
%	radiation pressure interaction in an optomechanical system. The slope with which the temporal dynamics of the mechanical oscillator, driven by radiation pressure interaction with an optical cavity illuminated by a blue--detuned laser, undergoes a Hopf bifurcation, and reaches a limit cycle, allows us to determine, in a simple and consistent way, the single--photon optomechanical rate, without the need of knowing the bath temperature. We provide the theoretical framework, and experimentally validate this method, providing a novel procedure for the full characterization of an optomechanical system, which could be extended outside cavity optomechanics, whenever a resonator is driven into a limit cycle by the appropriate interaction with another degree of freedom.
%\end{abstract}

\date{\today}

\maketitle

\section{Introduction}  
The dynamics of interacting radiation-matter systems crucially depend upon the value of their coupling rate compared to that of dissipative rates affecting the system, and typically very different physical phenomena occur in the weak, strong, or ultrastrong coupling regime. As a consequence, a reliable and precise method for the measurement of the coupling rate is a necessary tool for the full characterization of this class of systems. The way to measure the coupling rate depends upon the specific system and also upon the parameter regime. For example, in cavity QED, one can spectroscopically measure the vacuum coupling rate via the dressed level spacing only if the dissipation rates are small enough and one has a large enough atomic cooperativity \cite{Berman}. Cavity optomechanics \cite{rmp} is characterized by the dispersive, radiation-pressure-like interaction between mechanical and optical resonators and provides a relevant example of coupled radiation-matter systems.
In cavity optomechanics the main concern is the determination of the single--photon optomechanical coupling rate $g_0$~\cite{rmp,Gorodetsky:2010uq}, corresponding to the cavity frequency shift associated with the exchange of a single phonon, and which is often difficult to measure accurately \cite{Regal:2015}.
Different methods have been introduced for the determination of $g_0$.
i) In Ref.~\cite{Gorodetsky:2010uq} the integrated power spectral noise (PSN) around the frequency of a calibration tone, $\omega_b$, provides a ruler for the calibration of the PSN around the frequency of the mechanical mode under investigation, $\omega_m$. Hence $g_0$ is determined by assuming knowledge of the number of quanta at a given temperature, $\bar n_m$, and the modulation depth of the calibration tone, $\beta$, which can be measured independently.
The main issue with such a method is knowledge of $\bar n_m$, particularly at cryogenic temperatures \cite{Nielsen:2017}. ii) In unresolved sideband optomechanical systems, a sophisticated and involved alternative to infer $g_0$ is to perform optomechanically induced transparency (OMIT) measurements \cite{Rossi:2019}. In resolved sideband optomechanical systems, the OMIT measurement gets even more involved than the previous case \cite{Karuza:2013}. iii) Another technique is to extrapolate $g_0$ from the back--action--broadened linewidth once the intracavity photon number for a specified detuning is known~\cite{Regal:2013, Painter:2019, Painter:2020}.

Here we present a method for the determination of $g_0$, based on the non--linear temporal dynamics of the oscillator rather than its stationary frequency analysis.
The measurement scheme might be more accurate than that of Refs.~\cite{Gorodetsky:2010uq, Rossi:2019, Karuza:2013, Regal:2013, Painter:2019, Painter:2020} especially at low temperature, and it can be applied in either the weak or strong coupling regime. 
The present scheme, which does not need any calibration of the signal, compared to the other methods, 
exploits the fact that a mechanical oscillator driven by the radiation pressure interaction with an optical cavity illuminated by a blue--detuned laser, undergoes a Hopf bifurcation and reaches a limit cycle~\cite{Marquardt2006, Kippenberg2005, Carmon, Marquardt2006, Metzger, Krause:2015aa, Buks2019}. 
We show that the slope with which the amplitude of the displacement of the mechanical oscillator grows up in time allows to determine in a simple and consistent way $g_0$ without the need of knowing $\bar n_m$. Moreover, this slope is very robust to thermal noise. This method is very general and can be applied to a large class of physical systems, well outside cavity optomechanics, when a resonator ends up in a limit cycle due to the coupling with another degree of freedom, such as, for example, the cavity of a laser coupled to an active medium~\cite{Siegman1996}, or in nonlinear superconducting circuits~\cite{Josephson}.

\section{Theoretical analysis}
\begin{figure}[t!]
	\centering
	\includegraphics[width=0.9\linewidth]{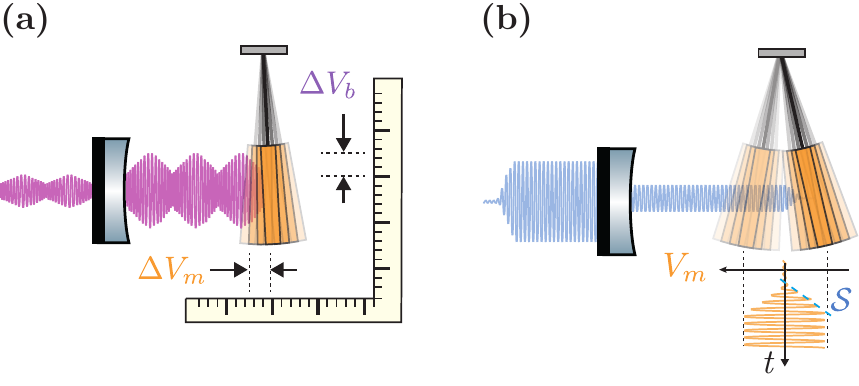}
	\caption{
		{\bf (a)} Determination of $g_0$ through the measurement of the stationary frequency noise fluctuations. Knowledge of the amplitude of the calibration tone, $\Delta V_b$, becomes a ruler for the fluctuations of the cavity resonance frequency, $\Delta V_m$.
		{\bf (b)}
		Determination of $g_0$ through the measurement of the slope $\calS$ of the temporal dynamics corresponding to reaching a limit cycle. The mechanical oscillator, which is excited by radiation pressure interaction with an optical cavity illuminated by a blue--detuned laser, undergoes a Hopf bifurcation, and reaches the limit cycle.}
	\label{fig:Figure_Sketch}
\end{figure}

The dynamics of a single mode of a mechanical oscillator, part of an optomechanical system, is usually revealed through coherent optical interrogation of a reflected or transmitted probe beam, while a pump beam is used for engineering the radiation pressure interaction. 
The method introduced in Ref.~\cite{Gorodetsky:2010uq} is based on the detection of the voltage spectral noise (VSN) of the probe beam around the mechanical frequency, $\omega_m$, and, at the same time, around the calibration tone, $\omega_b$, placed near the mechanical frequency for avoiding inevitable differences in the response of the opto--electronics devices. The single--photon optomechanical rate is found to be~\cite{Gorodetsky:2010uq}
\begin{equation}\label{eq:g0_freq}
	g_0 =
		\frac{1}{\sqrt{2 \bar n_m}}
		\frac{\beta\,\omega_b}{\sqrt{2}}
	\sqrt{\frac{ \Delta V^2_m}{ \Delta V^2_b}}\,
	\calK\,,
\end{equation}
where $\calK$ is a function dependent upon the detection system used, and which is given for our case in Appendix~\ref{App:A}.
Provided that the number of quanta at a given temperature, $\bar n_m$, is known, this expression requires knowledge of the modulation depth of the calibration tone $\beta$, which becomes a ruler for the fluctuations of the cavity resonance frequency [see Fig.~\ref{fig:Figure_Sketch}(a)].

We consider a method for the determination of $g_0$ by the measurement of the slope with which the mechanical oscillator, which is excited by the radiation pressure interaction with an optical cavity illuminated by a blue--detuned laser, reaches a limit cycle [see Fig.~\ref{fig:Figure_Sketch}(b)].
Qualitatively, the larger the optomechanical coupling $g_0$, the shorter the time the resonator takes to reach a stationary limit cycle.
%
% ----------------------------------------------
%
%-----------------------------------
%
%We can make more quantitative arguments using the commonly adopted slowly varying amplitude equations analysis~\cite{Marquardt2006,Holmes2012,Li:2020aa}, which is presented in Sup.~Mat., and which is valid whenever there is a clear timescale separation between the fast oscillations and the slow variation of the amplitude of the resonator. In cavity optomechanics this condition excludes only the ultra--strong coupling regime where $g_0$ becomes comparable to the mechanical frequency $\omega_m$.
%
%-----------------------------------
%
An optomechanical system constituted of two optical modes interacting via radiation--pressure with one mechanical mode might be described by the Hamiltonian
\begin{equation}\label{ham0_sup}
	H=H_{\rm pump}+H_{\rm probe}+H_{\rm mech}+H_{\rm int},
\end{equation}
decomposed as follows
\begin{eqnarray}
	H_{\rm pump} &=& \hbar \omega_{c,pm}a_{pm}^\dagger a_{pm}+\ii E_{pm}\left(a_{pm}^\dagger \ee^{-\ii\omega_{L,pm}t}- a_{pm} \ee^{\ii\omega_{L,pm}t}\right),\quad  \\
	H_{\rm probe} &=& \hbar \omega_{c,pr}a_{pr}^\dagger a_{pr}+\ii E_{pr}\left(a_{pr}^\dagger \ee^{-\ii\omega_{L,pr}t}- a_{pr} \ee^{\ii\omega_{L,pr}t}\right),  \\
	H_{\rm mech} &=& \hbar \omega_{m}b^\dagger b , \\
	H_{\rm int} &=& -\sum_{i=pr,pm}\hbar g_{i} a_i^\dagger a_i(b+b^\dagger).
\end{eqnarray}
We have a pump cavity mode with bosonic annihilation operator $a_{pm}$ and resonance frequency $\omega_{c,pm}$, which is driven at frequency $\omega_{L,pm}$. The probe mode, with resonance frequency $\omega_{c,pr}$ and providing a continuous, real-time detection of the mechanical motion, described by the bosonic annihilation operator $a_{pr}$, is driven in general at a different frequency $\omega_{L,pr}$, and it refers to a different cavity mode from that driven by the pump (different frequency and/or polarization) in order to avoid interference between the two drivings. The driving rates are explicitly given by $E_i=\sqrt{2\kappa_{in,i} P_i/\hbar\omega_{L,i}}$  [$i = (pr,pm)$] with $\kappa_{in,i}$ the i-th cavity mode decay rate through the input port, and $P_i$ the associated laser input power.
The mechanical Hamiltonian $H_{\rm mech}$ describes a membrane resonator with bosonic annihilation mechanical operator $b$, with resonance frequency $\omega_{m}$ and effective mass $m_{\rm eff}$, and $q = (b + b^\dag)\,x_{\rm zpf}$, where $x_{\rm zpf}=\sqrt{\hbar/2m_{\rm eff}\omega_{m}}$ is the spatial width of the oscillator zero point motion. Finally we have the usual radiation pressure dispersive interaction term between the optical pump and probe modes and the mechanical mode, quantified by the single-photon optomechanical coupling rates $g_{i}=-(d\omega_{c,i}/dx)x_{\rm zpf}$.
%where $m_j$ is the corresponding mass.

We then move to the interaction picture with respect to the optical Hamiltonian $H_0= \hbar \omega_{L,pr}a_{pr}^\dagger a_{pr}+\hbar \omega_{L,pm}a_{pm}^\dagger a_{pm}$, which means considering, for both pump and probe modes, the frame rotating at the corresponding laser driving frequency. The membrane resonator and the cavity modes are coupled to their corresponding thermal reservoir at temperature $T$ through fluctuation-dissipation processes, which we include in the Heisenberg picture by adding dissipative and noise terms. Moreover, we restrict our study non--linear dynamics at room temperature $T \simeq \SI{300}{\kelvin}$ only, which justifies a classical treatment of the Langevin equations, and implies a different treatment of optical and mechanical noise terms. In fact, at optical frequencies $\omega_L/2\pi \simeq 10^{14}$~Hz, so that the thermal excitation number is $\bar{n}_{L} \simeq 0$, while at mechanical frequencies $\omega_{m}/2\pi \simeq 10^{6}$ Hz implying $\bar{n}_m \simeq k_b T/\hbar\omega_{m} \gg 1$. As a consequence, we expect that thermal noise will be dominant for the mechanical mode, but for large enough driving powers, we cannot exclude in general the presence of non-negligible effects of the fluctuations of the intracavity field, due either to technical laser noise or ultimately to vacuum fluctuations.
Therefore, we consider classical complex random noises,  $\beta^{in}(t)$, and $\alpha_i^{opt}(t)$, with correlation functions
\begin{eqnarray}
	\label{corre1_sup}
	&\langle \beta^{in}(t) \beta^{in}(t')\rangle = \langle \alpha_i^{opt}(t) \alpha_{i'}^{opt}(t')\rangle = 0, \\
	&\langle \beta^{in,*}(t) \beta^{in}(t')\rangle =(\bar{n}_m+1/2) \delta(t-t'), \label{corre2_sup}\\
	&\langle \alpha_i^{opt,*}(t) \alpha_{i'}^{opt}(t')\rangle=(1/2)\delta_{i i'}\delta(t-t'), \label{corre3_sup}
\end{eqnarray}
% yielding the following quantum Langevin equations~\cite{Aspelmeyer2014,Giovannetti2001}, for $i,j=1,2$,
%
and $\langle \beta^{in}(t')\beta^{in,*}(t)\rangle=\langle \beta^{in,*}(t) \beta^{in}(t')\rangle$ and $\langle \alpha_i^{opt,*}(t) \alpha_{i'}^{opt}(t')\rangle=\langle \alpha_i^{opt}(t') \alpha_{i'}^{opt,*}(t)\rangle$ because the $c$-numbers lose the commutation relation. The quantum Langevin equations are therefore well approximated by the set of coupled classical Langevin equations for the corresponding optical and mechanical complex amplitudes $\alpha_i(t)$ and $\beta(t)$,
\begin{align}\label{eq:c_langevin1_sup}
	\dot{\alpha}_i(t)=&\left(\ii \Delta^{(0)}_i-\kappa_i\right)\alpha_i(t) \!+\! E_i +
	\nonumber\\
	&\qquad\qquad\,\,+\! 2\ii g_{i}\text{Re}[\beta(t)]\alpha_i(t)
	\!+\!\sqrt{2\kappa_i}\,\alpha_i^{opt}(t),
	\\
	\dot{\beta}(t)=&(-\ii \omega_{m}-\gamma_{m})\beta(t) +\!\! \sum_{i=pr,pm}\!\!  g_{i}\vert\alpha_i(t)\vert^2
	+\sqrt{2\gamma_{m}}\,\beta^{in}(t),
	\label{eq:c_langevin_sup}
\end{align}
where $\Delta^{(0)}_i=\omega_{L,i}-\omega_{c,i}$, $\kappa_i=\kappa_{in,i}+\kappa_{ex,i}$ is the total cavity amplitude decay rate for pump and probe modes, $\kappa_{ex,i}$ is the optical loss rate through all the ports different from the input one, and $\gamma_{m}$ is the amplitude decay rate of the membrane.

We can make more quantitative arguments using the commonly adopted slowly varying amplitude equations analysis~\cite{Marquardt2006,Holmes2012,Li:2020aa}, which is valid whenever there is a clear timescale separation between the fast oscillations and the slow variation of the amplitude of the resonator. In cavity optomechanics this condition excludes only the ultrastrong coupling regime where $g_0$ becomes comparable to the mechanical frequency $\omega_m$.
Discarding here the limiting case of chaotic motion of the resonator, which occurs however only at extremely large driving powers, which are not physically meaningful for the Fabry--P\'erot cavity system considered here, the mechanical resonator, after an initial transient regime, sets itself into dynamics of the form
\begin{equation}\label{ansatz_sup}
	\beta(t)=\beta_{0}+A(t)\ee^{-\ii \omega_{m} t},
\end{equation}
where $\beta_{0}$ is the approximately constant, static shift of the resonator, and $A(t)$ is the corresponding slowly-varying complex amplitude. Eq.~(\ref{ansatz_sup}) implies that we study the long--time dynamics of the mechanical resonator in the frame rotating at the fast mechanical frequency $\omega_{m}$.
Inserting Eq.~(\ref{ansatz_sup}) into Eq.~(\ref{eq:c_langevin1_sup}), solving it formally by neglecting the transient term related to the initial values $\alpha_i(0)$, expanding the intracavity field in terms of the Bessel function of the first kind, $J_n$, and neglecting the transient decay term, we arrive to the amplitude equation
\begin{eqnarray}
	\dot{A}(t)=-\gamma_{m}\, A(t)
			&+&\ii \,A(t) \sum_{i=pr,pm}g_i\,\mathcal{F}_i(\vert A(t)\vert ) + \nonumber \\
			&+&\ii\sum_{i=pr,pm}g_{i}\eta_i^{opt}(t) + \sqrt{2\gamma_{m}}\,\beta^{in}(t)\,, 
			\label{eq:first-order finnal0_sup}
\end{eqnarray}
where $\eta_i^{opt}(t)$ is an optical noise term depending upon the intracavity amplitude, and the non--linear contribution is
\begin{equation}
	\mathcal{F}_i=\frac{E_i^2}{\vert A\vert}
		\sum_{n=-\infty}^{\infty}
		\frac{J_n\left(-\xi_i\right)J_{n+1}\left(-\xi_i\right)}
		{[\ii n\omega_{m}-\calL_i][-\ii (n+1)\omega_{m}-\calL_i^*]},
		\label{eq:auxiliary function_sup}
\end{equation}
which, in turn, depends upon the corresponding variable $\xi_i = 2g_i\vert A\vert/\omega_{m}$, and $\calL_i=\ii  \Delta_i-\kappa_i$, where $\Delta_i = \Delta^{(0)}_i+ (\beta_{0}+\beta^{*}_{0})g_{i}$ is the {\it effective detuning}, that is the detuning between the laser and the cavity resonance frequencies modified by the static optomechanical interaction. 
For further details on these calculations, we refer the reader to Ref.~\cite{Piergentili2020}. As already shown in Refs.~\cite{Holmes2012,Li:2020aa}, Eq.~(\ref{eq:first-order finnal0_sup}) provides a general and very accurate description of the dynamics of the mechanical resonator.

To study the regime where the oscillator reaches a limit cycle, one can make quantitative predictions on such a regime assuming that $|A| \gg \sqrt{2 \bar{n}_m}$ neglecting therefore mechanical thermal noise. Moreover, in our experiment the effect of optical noise is negligible and we do not consider the terms associated with $\eta_i^{opt}(t)$.
With the above approximations, Eq.~(\ref{eq:first-order finnal0_sup}) becomes
\begin{eqnarray}\label{eqampliappr1_sup}
	\dot{A}(t)	\!=\! -\Big[\gamma_{m}^{eff}(|A|) - \ii \Delta\omega_{m}^{eff}(|A|)\Big]\, A(t),
\end{eqnarray}
where, assuming that $g_{i} \simeq g_0$, the effective detuning can be cast as
\begin{eqnarray}
	\Delta\omega_{m}^{eff}(|A|) 
	=  \frac{g_0}{|A|}\text{Re}\left[E_{pr}^2\,\Sigma_{pr}(|A|) + E_{pm}^2\,\Sigma_{pm}(|A|)\right]\,,
	\label{eq:first-order delta finnal_sup}
\end{eqnarray}
and the effective mechanical damping as 
\begin{eqnarray}
	\gamma_{m}^{eff}(|A|) = \gamma_{m}
	\left[
	1\! +\!  \frac{g_0}{\gamma_{m}|A|}\text{Im}\left[E_{pr}^2\,\Sigma_{pr}(|A|) + E_{pm}^2\,\Sigma_{pm}(|A|)\right]
	\right]\!,\quad\,\,\,\,
	\label{eq:first-order finnal_sup}
\end{eqnarray}
where 
\begin{eqnarray}
	\Sigma_i(|A|) = \sum_n\frac{J_n\left(-\xi\right)J_{n+1}\left(-\xi\right)}
	{[\ii n{\omega_{m}}-\calL_i][-\ii (n+1)\omega_{m}-\calL_i^*]}\,.
	\label{eq:Sigmaj_sup}
\end{eqnarray}
with $\xi = 2g_0\vert A\vert/\omega_{m} =  g_0 q/\omega_{m}x_{\rm zpf}$.
Eq.~(\ref{eqampliappr1_sup}) can be solved by rewriting it in terms of the modulus and phase, $A = I \ee^{\ii \phi}$,
\begin{eqnarray}\label{eqamplimodu_sup}
	\dot{I}(t)&=&-\gamma_{m}^{eff}(I)\, I(t)\, \label{eqamp_sup}\\
	\dot{\phi}(t)&=& \Delta\omega_{m}^{eff}(I)\,, \label{eqfase_sup}
\end{eqnarray}
Finally, by using the relation  $I = \vert A\vert = \xi\,\omega_{m}/2g_0$, 
%eq.~\eqref{eqamp_sup}
% can be cast as in eq.~(2) of the Letter ($1\rightarrow pm$, and $2\rightarrow pr$)
%%
%\begin{eqnarray}\label{final_sup}
%	\dot{\xi}=-\gamma_{\rm m }\left[\xi + \frac{2g_0^2}{\gamma_{m}{\omega_{m}}}\text{Im}\left[E_1^2\,\Sigma_1 + E_{pm}^2\,\Sigma_{pm}\right]  \right]
%	\,.
%\end{eqnarray}
%
%-----------------------------------
%
%% ----------------------------------------------
%
%The 
the dynamics of the slowly varying mechanical displacement amplitude $q$ can be cast using the dimensionless quantities $\xi$ and $\tau = t \gamma_m$, as
\begin{eqnarray}\label{eq:xi}
  \dot{\xi}(t)&=& - \calS(\xi)\,,
\end{eqnarray}
where the derivatives, $\calS$ and $\Sigma^\prime$,  are taken with respect to $\tau$, and the
slope function $\calS(\xi)$ is~\cite{Piergentili2020}
\begin{eqnarray}
	\calS(\xi) =\xi +  \alpha\,\text{Im}\Big[E^2_{pm}\Sigma_{pm}(\xi) +  E^2_{pr}\Sigma_{pr}(\xi) \Big]\,,
	\label{eq:first-order finnal}
\end{eqnarray}
with $\alpha = 2g_0^2/\gamma_m\omega_m$.
%$E_i = \sqrt{2\kappa_{in,i} P_i/\hbar\omega_L}$, and
%%
%\begin{eqnarray}\label{eq:Sigmaj}
%	\Sigma_i(\xi)  = \sum_n\frac{J_n\left(-\xi\right)J_{n+1}\left(-\xi\right)}
%					{[\ii n{\omega_m}-\calL_{i}][-\ii (n+1)\omega_m-\mathcal{L}^\ast_{i}]}
%					\nonumber \,,
%\end{eqnarray}
%%
%where $J_n$ is the $n-$th Bessel function, $\calL_j = \ii \Delta_j - \kappa_j$ is evaluated for the driven cavity modes associated with the probe and pump beam, respectively [$i = (pr,pm)$].
It is worth noting that the contributions in Eq.~(\ref{eq:first-order finnal}) are null for $\Delta_{i} = 0$, that is, perfectly resonant beams do not introduce optomechanical effects (see Appendix~\ref{App:B}).
After a transient, Eq.~(\ref{eq:xi}) yields a stationary steady state with a constant radius of the limit cycle, $q^{st}$, equal to a strictly positive root of the implicit equation
%for $\xi \neq 0$, as
%
\begin{equation}\label{eq:slopeeff}
  	\calS(\xi^{st}) = \xi^{st} +  \alpha\,\text{Im}\Big[E^2_{pm}\Sigma_{pm}(\xi^{st}) +  E^2_{pr}\Sigma_{pr}(\xi^{st}) \Big]= 0\,.
\end{equation}
Because of the oscillating behavior of the Bessel functions, Eq.~(\ref{eq:slopeeff}) may have more than one stable solution (see e.g., Ref.~\cite{Marquardt2006}), but usually the resonator starts from a very small amplitude of thermal origin and the stationary dimensionless amplitude $\xi^{st}$ corresponds to the \textit{smallest} positive root of Eq.~(\ref{eq:slopeeff}).
We stress that the occurrence of more than one stable solution for Eq.~\eqref{eq:slopeeff} takes place at large enough driving power~\cite{Marquardt2006}, and our method, instead, applies at the onset of the Hopf bifurcation, that is, at low driving power.

A first important observation is that the excitation process with which the mechanical oscillator reaches the limit cycle depends on $g_0$. The extreme of the slope function $\calS(\xi)$, for given input powers, is provided by
\begin{eqnarray}\label{eq:dslope}
  	\calS^\prime(\xi^{mx}) = 1 + \alpha\,\text{Im}\Big[E^2_{pm}\Sigma^\prime_{pm}(\xi^{mx})
									+  E^2_{pr}\Sigma^\prime_{pr}(\xi^{mx}) \Big] = 0,\quad
\end{eqnarray}
where the derivative is taken with respect to $\xi$, which allows us to determine the amplitude $\xi^{mx}$. As a consequence, 
the maximum slope is given by
\begin{equation}\label{eq:dslopemax}
  \calS(\xi^{mx}) = \xi^{mx} +  \alpha\,\text{Im}\Big[E^2_{pm}\Sigma_{pm}(\xi^{mx}) +  E^2_{pr}\Sigma_{pr}(\xi^{mx}) \Big],
\end{equation}
which is an increasing function of both $g_0$ and the input powers. 
Inspection of Eq.~\eqref{eq:dslopemax} shows that probe and pump beams are theoretically equivalent, and in the absence of a pump beam even a slightly blue--detuned  ($\Delta_{pr}>0$) probe beam, always necessary for locking the laser and the cavity resonance frequencies, is sufficient for our purpose. However, in the following we limit ourselves to the case of a perfectly resonant probe beam ($\Delta_{pr} = 0$), which allows us to take $\Sigma_{pr}(\xi)=0$ in the above equations, and focus only on the effect provided by the presence of the blue--detuned ($\Delta_{pm}\!\sim\!\omega_m$) pump beam.
The second important observation is that it is possible to define a {\it threshold pump power} $P_{pm}^{th}$, as the minimum power for which Eq.~\eqref{eq:slopeeff} is satisfied for $\xi \ne 0$, and which allows an estimation of $g_0$ only knowing $P_{pm}^{th}$. In fact, as the pump power approaches such threshold %\blu{from above}
, both the amplitude $\xi^{mx}$ and the maximum slope $\calS(\xi^{mx})$ tend to zero, and from Eq.~\eqref{eq:dslope}, or, equivalently, Eq.~\eqref{eq:dslopemax}, one has the constraint, for $\Delta_{pr} = 0$,
\begin{eqnarray}\label{eq:Fmx}
  	\calA  \equiv 
  		\lim_{\xi \rightarrow 0 }
  		\frac{-1}{\text{Im}\big[E^2_{pm}\Sigma^\prime_{pm}(\xi)\big]} 
  	= 
  		\lim_{\xi \rightarrow 0 }\frac{-\xi}{\text{Im}\big[E^2_{pm}\Sigma_{pm}(\xi) \big]} = \alpha
  	\,,
\end{eqnarray}
so that,
\begin{equation}\label{eq:Pthrmax}
  	g_0^2 = \gamma_m\omega_m\,
			\frac{\calA}{2}
			%\blu{\,,}
\end{equation}
%
%\blu{where, using Eq. (18) and the second expression in Eq. (26), 
%\begin{align}
%	{\cal A}=\frac{\hbar \omega_L}{\kappa_{in}P^{th}}\frac{[\kappa^2+\Delta^2] [(\kappa^2+\Delta^2-\omega_m^2)^2+4\kappa^2 \omega_m^2]}{4\omega_m^2 \Delta \, \kappa},
%\end{align}
%}
is a function only of known and experimentally determinable parameters.%\blu{, and we have omitted the index $pm$ for simplicity in this explicit expression.}

\section{Experimental validation}
\begin{figure}[h!]
	\centering
	\includegraphics[width=0.9\linewidth]{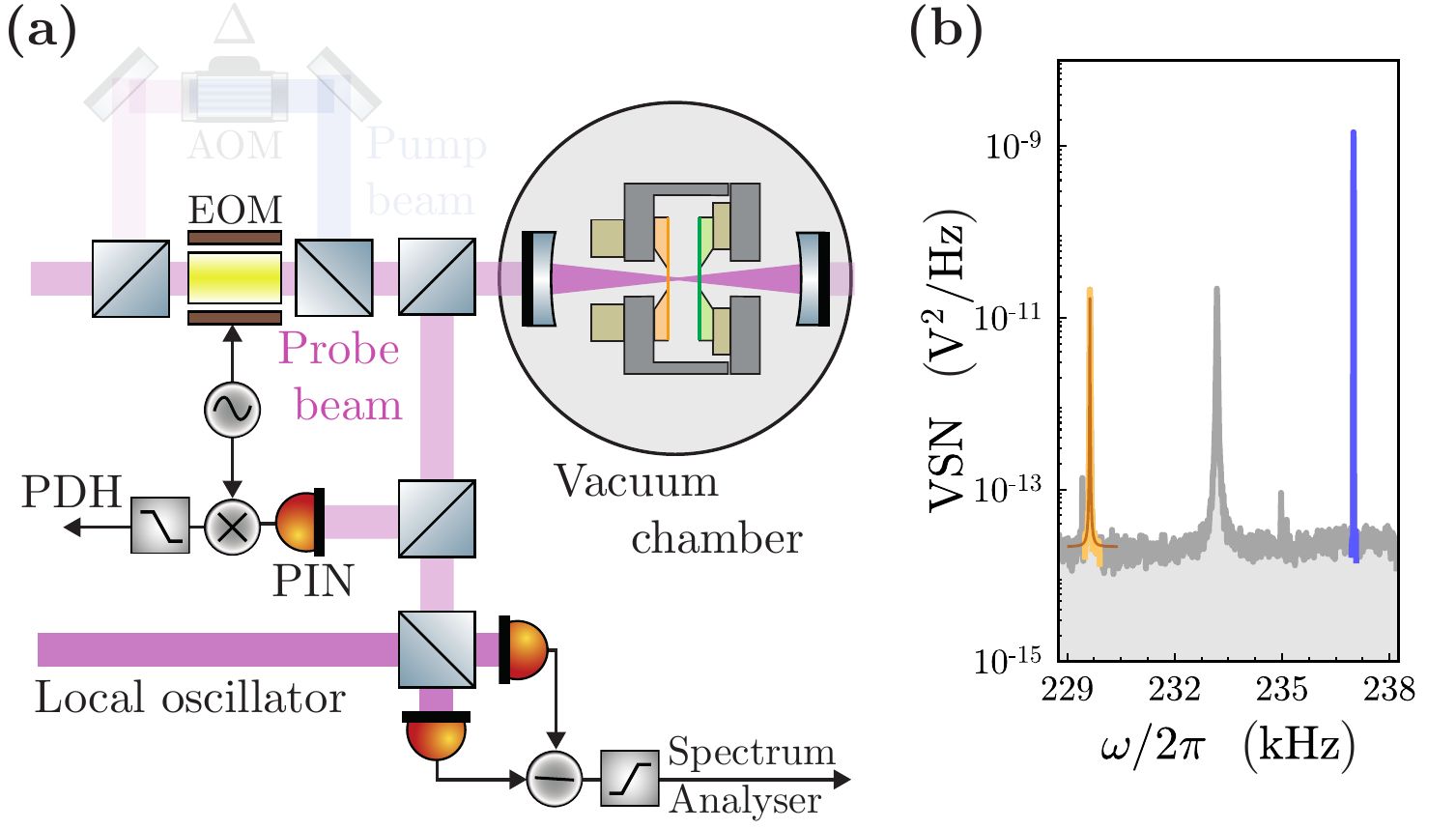}
	\caption{
	{\bf (a)} Experimental setup. A probe beam, frequency modulated by an EOM, impinges on the optical cavity. The reflected beam is analyzed by homodyne detection in order to detect the mechanical motion.
	{\bf (b)} VSN of the reflected field around the fundamental mode of the membrane.
	The orange area indicates the fundamental mode of the membrane under consideration; the blue area represents the calibration tone implemented by modulating the input field by means of an EOM; the peak in the middle represents the fundamental mode of the second disregarded membrane.
 }
	\label{fig:Figure_Freq}
\end{figure}

We validate experimentally our idea by means of the optomechanical system described in detail in Ref.~\cite{Piergentili:2018aa,Piergentili2020}, constituted by a two--membrane sandwich placed in the middle of an optical cavity [see Fig.~\ref{fig:Figure_Freq}(a)]. However, the second membrane, under proper experimental conditions, remains essentially in its thermal state and can be ignored in the following~\cite{Piergentili2020}.

The membrane we use in the experiment is a high--stress $\mathrm{Si_3N_4}$ square membrane produced by Norcada, with a side of \SI{1.5}{\milli\meter}, and a thickness of $\SI{106}{\nano\meter}$. The membrane is placed inside a \SI{90}{\milli\meter}--length optical cavity with empty cavity finesse $\calF_0 = \num{50125\pm 25}$ \cite{Rossi:2017aa}, and the optomechanical system is located in a vacuum chamber evacuated to \SI{5e-7}{\milli\bar}. The probe beam is locked to the optical cavity by means of the Pound--Drever--Hall technique, and the membrane displacement is unveiled by homodyne detection of the light reflected by the optical cavity.

Firstly, we consider the determination of $g_0$ by means of Eq.~\eqref{eq:g0_freq}. A typical VSN measurement is reported in Fig.~\ref{fig:Figure_Freq}(b). The orange area indicates the voltage noise of the fundamental mode of the oscillator we consider for describing our idea, for which we estimate $\omega_{m} = 2\pi \times\SI{229.753\pm0,0001}{\kilo\hertz}$, $\gamma_m = 2\pi \times\SI{1.64\pm0.1}{\hertz}$. The blue peak is the calibration tone at frequency $\omega_b = 2\pi \times\SI{237.0}{\kilo\hertz}$, and it is implemented by modulating the input field by means of an electro--optical modulator (EOM), and calibrated by heterodyne detection of the beating signal between carrier and sidebands (see Appendix~\ref{App:A1}).
From Fig.~\ref{fig:Figure_Freq}(b), the measured voltage noises are $\Delta V^2_m  = \SI{2.8\pm0.4 e-10}{\square\volt}$, and $\Delta V^2_b  = \SI{1.669\pm0.001 e-8}{\square\volt}$ (see Appendix~\ref{App:A2}).
Finally, by considering the values of the parameters $\beta = \SI{19.5\pm0.1}{\milli\radian}$, $\bar n_m = \num{267.6\pm0.5 e5}$ for the measured room temperature $T =\SI{295\pm0.5}{\kelvin}$, and $\calK = \num{5.65\pm0.14}$, whose determination is detailed in Appendix~\ref{App:A}, we estimate $g_0 = 2\pi\times \SI{0.327\pm0.033}{\hertz}$, where the larger contribution to the standard deviation is due to the estimation of the integral of the VSN.

\begin{figure}[b!]
	\centering
	\includegraphics[width=.9\linewidth]{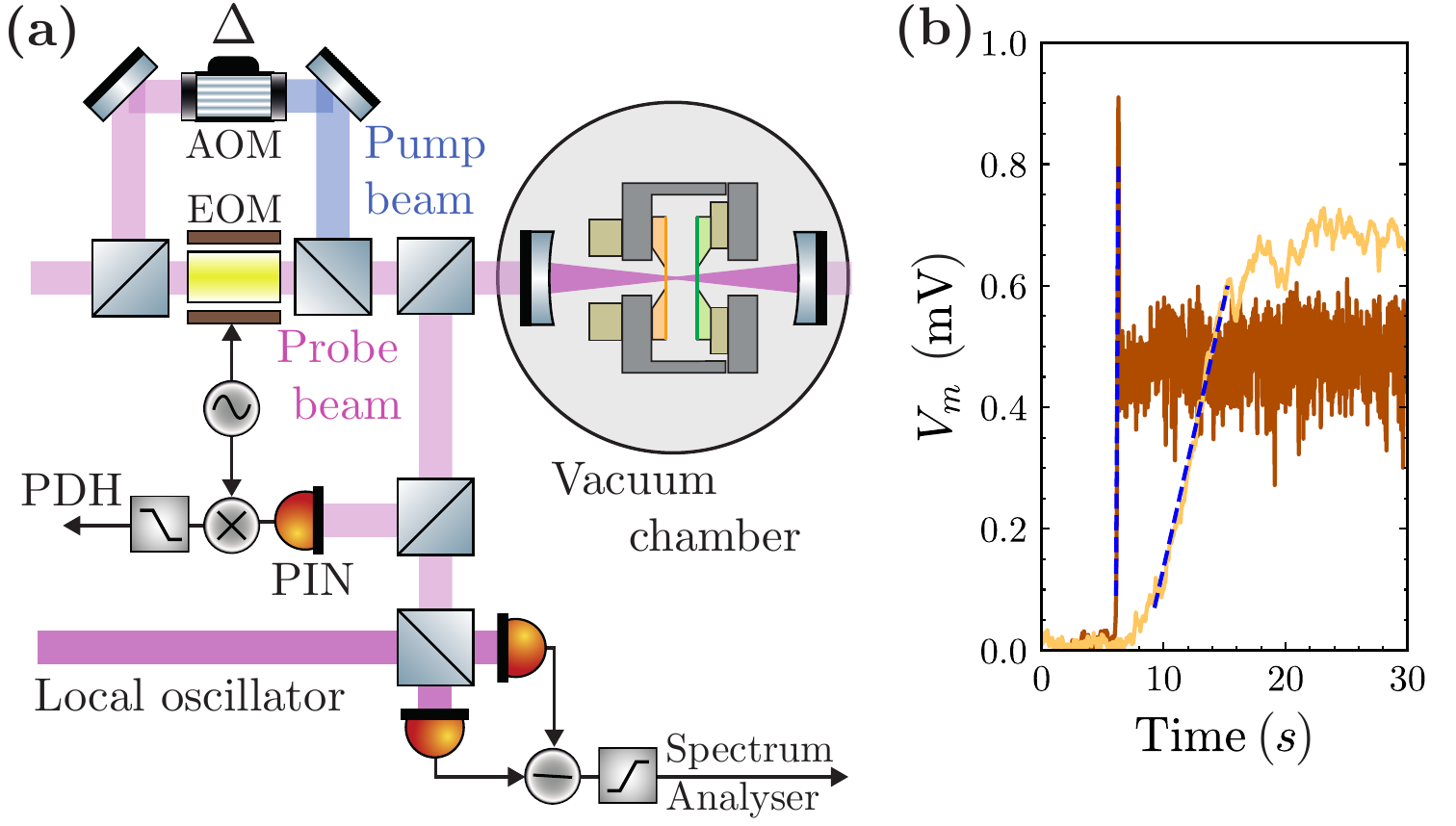}
	\caption{
	{\bf (a)} Experimental setup. The pump beam, blue--detuned by $\Delta = \SI{239.35}{\kilo\hertz}$ from the cavity resonance, is added to the experimental setup of Fig.~\ref{fig:Figure_Freq} for engineering the optomechanical interaction. 
	{\bf (b)} Voltage noise $V_m$ acquired around $\SI{229.753}{\kilo\hertz}$ in a bandwidth of $\SI{182}{\hertz}$, as a function of time. The time trace in the first \SI{5}{\second} corresponds to the thermal dynamics of the mechanical oscillator. After \SI{5}{\second} the pump beam is turned on, and the amplitude of the mechanical oscillator rises to a limit cycle. The orange and brown curves correspond to input power $P_{pm} = \SI{6.1\pm0.2}{\micro\watt}$ and $P_{pm} = \SI{21.0\pm0.5}{\micro\watt}$, respectively.
	}
	\label{fig:Figure_Time}
\end{figure}

We now consider the determination of $g_0$ by means of Eq.~\eqref{eq:dslopemax}, and the more direct estimation based on Eq.~\eqref{eq:Pthrmax}. The method consists in estimating the slope of the output signal for the rising of the amplitude of a mechanical oscillator from thermal state to a limit cycle due to the blue--detuned pump beam [see Fig.~\ref{fig:Figure_Time}(a)].
The acquired homodyne signal is demodulated around $\SI{229.753}{\kilo\hertz}$ in a bandwidth of $\SI{182}{\hertz}$, by means of a lock--in amplifier, as a function of time. The phase and quadrature components, $V_x$ and $V_y$, respectively, allow us to determine the amplitude on the signal $V_m = \big(V_x^2 + V_y^2\big)^{1/2}$. In Fig.~\ref{fig:Figure_Time}(b) are reported two time traces obtained for two different pump powers, $P_{pm} = \SI{6.1\pm0.2}{\micro\watt}$ and $P_{pm} = \SI{21.0\pm0.5}{\micro\watt}$, orange and brown curves, respectively.
The first $\SI{5}{\second}$ indicates the thermal displacement of the mechanical oscillator, while, after turning on the pump beam, the displacement amplitude of the oscillator increases, and reaches a stable amplitude (limit cycle for the mechanical oscillator) in a time that depends strongly on the pump power: the orange curve in Fig.~\ref{fig:Figure_Time}(b), corresponding to lower pump power, shows a much slower increase than the brown curve, which corresponds to higher power, for the same radiation pressure interaction rate $g_0$. This is the effect that allows us to determine $g_0$.
\begin{figure}[t!]
\begin{center}
   {\includegraphics[width=.35\textwidth]{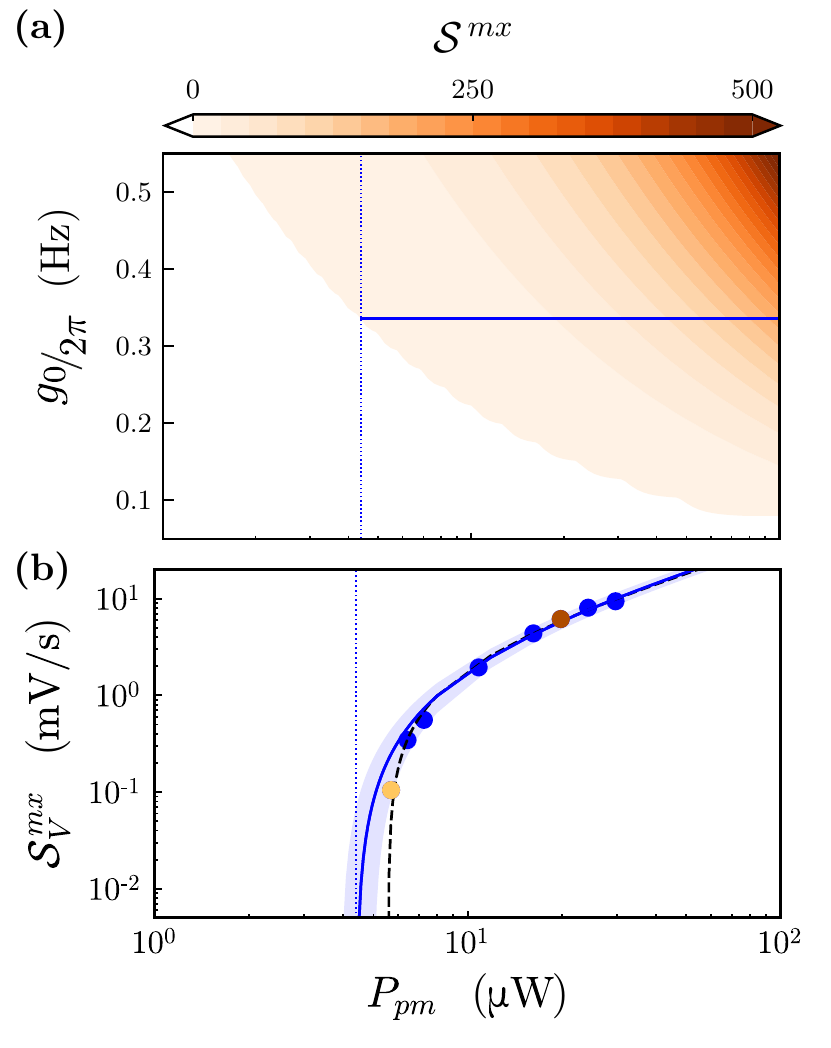}}
 \caption{
 	{\bf (a)} Dimensionless maximum slope, $\calS^{mx} = \calS(\xi^{mx}) $, as a function of the pump power, $P_{pm}$, and the single--photon optomechanical coupling rate $g_0$, evaluated for our experimental parameters.
	The blue line indicates the maximum slope for $g_0 = 2\pi\times\SI{0.336}{\hertz}$, and the dashed--blue line the threshold pump power $P_{pm}^{th} \simeq \SI{4.4}{\micro\watt}$, obtained for the experimental results presented in panel (b).
	{\bf (b)} Maximum slope multiplied by a transduction constant coefficient $a~[\si{\volt\per\second}]$, $\calS_V^{mx} =  a\,\calS^{mx}$, as a function of the pump power, $P_{pm}$.
	Filled circles represent experimental data; orange and brown circles correspond to the time traces reported in Fig.~\ref{fig:Figure_Time}(b).
	The blue curve represents the best--fit result of eq.~\eqref{eq:dslopemax} by using the Levenberg--Marquardt fitting method with fitting parameters $g_0$, and the transduction constant coefficient $a$; the best--fit values are given with $1-\sigma$ ($\sim 68\%$ confidence) uncertainty (light--blue area).
 	The dashed--black curve represents a linear fit for the determination of the threshold pump power.
}
\label{fig:Slope_LOG}
\end{center}
\end{figure}
In particular, as shown in Fig.~\ref{fig:Figure_Time}(b), we determine the rate at which the voltage noise, initially due to a thermal displacement of the mechanical oscillator, reaches a limit cycle, for different input pump powers $P_{pm}$. Such data, reported in Fig.~\ref{fig:Slope_LOG}(b) (and presented in Appendix~\ref{App:C} calibrated in terms of displacement units), are fitted with the expression of the dimensionless maximum slope $\calS^{mx} = \calS({\xi^{mx}})$, given by Eq.~\eqref{eq:dslopemax},  rescaled by a transduction constant coefficient $a~[\si{\volt\per\second}]$, $\calS_V^{mx} =  a\,\calS^{mx}$.
The blue curve in Fig.~\ref{fig:Slope_LOG}(b) represents the best--fit result for the maximum slope in Eq.~\eqref{eq:dslopemax} by using the Levenberg--Marquardt fitting method, with fitting parameters $g_0$, and $a$, only, resulting in the best--fit values $a= \SI{2.1 \pm 0.3 e-04}{\volt\per\second}$, and $g_0 = 2\pi\times\SI{0.336\pm0.013}{\hertz}$. This value of $g_0$ is consistent with that obtained by the calibration tone technique. Furthermore, it is given with a smaller standard deviation, which is limited by the accuracy with which the slope and input power are estimated.
	
A faster and simpler way for the determination of $g_0$ is provided by Eq.~\eqref{eq:Pthrmax}, with an estimation of the threshold pump power $P_{pm}^{th}$, given either by increasing it until the amplitude of the mechanical oscillator starts to rise, or by a linear fit of the data. In Fig.~\ref{fig:Slope_LOG}(b) the dashed--black curve represents a linear fit for the determination of the best--fit value of the threshold pump power $P_{pm}^{th} = \SI{5.6\pm0.7}{\micro\watt}$, consistent with the minimum experimental pump power able to establish a limit cycle [orange curve in Fig.~\ref{fig:Figure_Time}(b), and orange circle in Fig.~\ref{fig:Slope_LOG}(b)]. 
%
%\blu{In our case, we infer from Eq.~\eqref{eq:Pthrmax}, %evaluating ${\calA \simeq \num{1.72e6}}$, evaluating $\calA$ at the power of \SI{1}{\micro\watt}, $\calA_{\mu W} \simeq \num{2.64e-6}$,%{\radian\per\second^2}}$,
%
In our case, we infer from Eq.~\eqref{eq:Pthrmax}, evaluating $\calA \simeq \SI{2.64e-12}{\watt}/P_{pm}^{th}(\si{\watt})$,
%
%\begin{eqnarray}\label{eq:A}
% 	g_0^2=
%	\frac{\gamma_m\omega_m}{P_{pm}^{th}}\frac{h c}{\lambda_0\,2\kappa_{in}}
%	\frac{\calA}{2}
%	\simeq (2\pi)^2
%	\frac{\SI{0.497}{\square\hertz\micro\watt}}{P_{pm}^{th}(\si{\micro\watt})}
%	\,,
%\end{eqnarray}
%\begin{eqnarray}\label{eq:A}
% 	g_0^2=
% 	\gamma_m\omega_m
% 	\frac{\calA_{\mu W}}{2}
% 	\frac{\SI{1}{\micro\watt}}{P_{pm}^{th}(\si{\watt})}
% 	\simeq (2\pi)^2
% 	\frac{0.498\,(\si{\square\hertz\micro\watt})}{P_{pm}^{th}\,(\si{\micro\watt})}
%	\,,
%\end{eqnarray}
\begin{eqnarray}\label{eq:A}
	g_0^2=
	\frac{\omega_m\gamma_m}{2}
	\frac{\SI{2.64e-12}{\watt}}{P_{pm}^{th}(\si{\watt})}
	\simeq (2\pi)^2
	\frac{0.497\,\si{\square\hertz\micro\watt}}{P_{pm}^{th}\,(\si{\micro\watt})}
	\,,
\end{eqnarray}
%
%where $\lambda_0$ is the laser wavelength, $\kappa_{in}$ is the decay rate of the probe beam, and $P_{pm}^{th}$ must be provided in $\si{\micro\watt}$.
where $P_{pm}^{th}(\si{\micro\watt})$ must be provided in microwatts.
Given $P_{pm}^{th} = \SI{5.6\pm0.7}{\micro\watt}$, we estimate $g_0 \simeq 2\pi\times \SI{0.705}{\hertz\sqrt{\micro\watt}}/\sqrt{\SI{5.6}{\micro\watt}}  \simeq2\pi\times \SI{0.298}{\hertz}$, which is close, but lower than the previously derived value, due to the overestimation of $P_{pm}^{th}$ by the linear fit.
Considering instead the intercept of the blue curve, one can more accurately estimate $P_{pm}^{th} \simeq \SI{4.4}{\micro\watt}$, yielding a more accurate value $g_0 \simeq 2\pi\times\SI{0.336}{\hertz}$, which coincides with the fitting procedure described above.

\section{Conclusion}
We have theoretically proposed and experimentally demonstrated a method for the accurate measurement of the single-photon optomechanical coupling rate $g_0$. The mechanical resonator when it undergoes a Hopf bifurcation, and is driven into a limit cycle by the radiation-pressure-like interaction, can be described in terms of a slowly varying amplitude equation, which predicts the growth of the amplitude from negligible values to the nonzero stationary amplitude of the limit cycle. The slope of such an amplitude increase can be expressed in simple terms as a function of $g_0$ and the driving power. One can derive with a best-fit analysis an accurate value of $g_0$, measuring the maximum slope from the time traces at different values of the input power. The only requirement is knowledge of the power characterizing the Hopf bifurcation and the birth of a limit cycle, the damping rate and the resonance frequency of the mechanical resonator, and the cavity coupling rates. This is in contrast with other methods that require either the additional knowledge of the temperature (or the mean thermal excitation number), or a more involved measurement. Interestingly, the method could be applied for the direct measurement of radiation-matter coupling rates outside cavity optomechanics, for example in laser systems where the cavity is excited to a limit cycle by the interaction with the active medium.
\begin{acknowledgments}
We acknowledge the support of the European Union Horizon 2020 Programme for Research and Innovation through the Project No. 732894 (FET Proactive HOT) and the Project QuaSeRT funded by the QuantERA ERA-NET Cofund in Quantum Technologies. P. Piergentili acknowledges support from the European Union's Horizon 2020 Programme for Research and Innovation under grant agreement No. 722923 (Marie Curie ETN - OMT).
\end{acknowledgments}

\appendix 
\section{Evaluation of $g_0$ by means of VSN}
\label{App:A}

The homodyne error signal detected in the setup of Figs.~\ref{fig:Figure_Freq}(a) and~\ref{fig:Figure_Time}(a), through coherent optical interrogation of a reflected probe beam, assuming optimal detection, is given by
\begin{eqnarray}
	V_H(t) = 2g_TS\sqrt{\mathcal{P}_{lo}\mathcal{P}_{in}}\,\,\text{Im}\left[\calR \right]\,,
\end{eqnarray}
where $\calR$ is the cavity reflection function
\begin{eqnarray}\label{eq:R0_sup}
	\calR
	 =  \sum_{b}\!J_b(-\beta)\,\ee^{\ii b\omega_bt}
	 \!\left[
	-1 +\ 2\kappa_{in}\!\!\sum_{l,n}\!
	\frac{J_{l-n}(-\xi)J_l(-\xi)\,\ee^{\ii n\omega_{m}t}}{\ii(l\omega_{m} +b\omega_b- \Delta) + \kappa}
	 \!\right]\,\,\quad
	\,,
\end{eqnarray}
with $\xi = g_0 q/\omega_m$, $\Delta = \omega_L - \omega_c$ the detuning of the laser frequency, $\omega_L$, with respect to the cavity resonance frequency, $\omega_c$, $\kappa$ the amplitude cavity decay rate, $S(\si{\ampere\per\watt})$ the sensitivity of the photodiodes, $g_T(\si{\volt\per\ampere})$ the transimpedance gain, $\mathcal{P}_{in}$ the input power, and $\mathcal{P}_{lo}$ the local oscillator power.
This voltage contains the signature of any modulation frequency of the reflected field, provided that it falls within the bandwidth of the electronic system.
The single--sided  power spectral density is
\begin{eqnarray}\label{eq:VHOmega_sup}
	 \calS_{W}(\Omega) = \frac{\calS_0}{4}
	 \left | \calR_+(\Omega) - \calR_-^\ast(\Omega) \right |^2
	\,,
\end{eqnarray}
with $ \calS_0 = (2g_TS)^2\,\mathcal{P}_{lo}\mathcal{P}_{in}/R_0$, where $R_0$ represents the termination resistor over which the power spectral noise is revealed.

We are interested at the noise contribution around the mechanical mode at $\omega_m$, which, for small modulation of the input field, i.e., $\beta\ll1$, is given by 
\begin{eqnarray}\label{eq:ImRm1_sup}
	\calR_\pm(\omega_{m}) \simeq J_0(-\beta)\,
		2\kappa_{in}\sum_{n}\frac{J_{n}(-\xi)J_{n\mp1}(-\xi)}{\ii (n\omega_{m} - \Delta) + \kappa}
			\,,
\end{eqnarray}
and calibration tone at  $\omega_{b}$,
\begin{eqnarray}\label{eq:ImRb_sup}
	\calR_\pm(\omega_b) \simeq J_{\pm 1}(-\beta)
		\left[
		-1 + 2\kappa_{in}
		\sum_{n}\frac{J^2_{n}(-\xi)}{\ii (n\omega_{m} \pm \omega_b- \Delta) + \kappa}
		\right]\quad
	\,.
\end{eqnarray}
For small displacement of the mechanical oscillator, i.e., $\xi \ll1$, 
\begin{eqnarray}\label{eq:ImRm1_sup}
	\calR_\pm(\omega_{m}) \simeq J_0(-\beta)\,
		J_0(-\xi)J_1(-\xi)
		\frac{2\kappa_{in}}{\kappa - \ii \Delta}\,
		\frac{\mp\ii \omega_m}{\ii \omega_{m} \pm(\kappa-\ii \Delta)}
			\,,\,\,\quad
\end{eqnarray}
and
\begin{eqnarray}\label{eq:ImRb_sup}
	\calR_\pm(\omega_b) \simeq J_{\pm 1}(\beta)
		\left[
		-1 + 2\kappa_{in}
		\frac{J^2_{0}(-\xi)}{\pm\ii\omega_b  + \kappa-\ii  \Delta}
		\right]
	\,.
\end{eqnarray}
The ratio of the integrated power spectral density, $\Delta V^2 = \int  \calS_{W}(\Omega) d\Omega R_0$, around those frequencies, $\omega_m$ and $\omega_b$, assuming that $\Delta = 0$, $J_0(x)\simeq 1$, and $J_1(x)\simeq x/2$, and considering that, for thermal displacement, $q = \sqrt{2\langle n_m\rangle}$, provides
\begin{eqnarray}
	\sqrt{\frac{ \Delta V^2_m}{ \Delta V^2_b}}\simeq 
	\frac{g_0\sqrt{2 \bar n_m}}{\beta\omega_b}
	\frac{2\kappa_{in}/\kappa}{\sqrt{1 + (\kappa-2\kappa_{in})^2/\omega_b^2}}
	\sqrt{\frac{\kappa^2 + \omega_b^2}{\kappa^2 + \omega_m^2}}	
	\,.\quad
\end{eqnarray}
Finally we obtain the estimation of $g_0$ knowing the modulation depth $\beta$ as
\begin{equation}
	g_0 \simeq 
		\frac{1}{\sqrt{2 \bar n_m}}
		\frac{\beta\omega_b}{\sqrt{2}}
	\sqrt{\frac{ \Delta V^2_m}{ \Delta V^2_b}}\,
	\calK	\,,
\end{equation}
which represents the expression for evaluating the single--photon optomechanical coupling rate $g_0$, where 
\begin{equation}
	\calK = \sqrt{2}\frac{\kappa}{2\kappa_{in}}
	\sqrt{1 + \frac{(\kappa-2\kappa_{in})^2}{\omega_b^2}}
	\sqrt{\frac{\kappa^2 + \omega_m^2}{\kappa^2 + \omega_b^2}}	
	\,.
\end{equation}
We stress that the function $\calK$ defines our specific detection system, and it is different from the expression  in Ref.~\cite{Gorodetsky:2010uq}  due to a different choice of local oscillator. In fact in Ref.~\cite{Gorodetsky:2010uq} the local oscillator is modulated in phase according to the input field. With such a choice, the calibration tone and the mechanical displacement {\it produce the same transduction coefficient in terms of spectral densities}~\cite{Gorodetsky:2010uq}, that is $\calK = 1$. 

In our setup the cavity length is $\SI{90\pm1}{\milli\meter}$, which gives a ${\rm FSR} = 2\pi\times\SI{1.67\pm0.02}{\giga\hertz}$.
The empty cavity finesse, estimated by ring--down techniques~\cite{Rossi:2017aa}, is $\mathcal{F}_0 = \num{50125\pm25}$. For a cavity realized with identical mirrors,  we infer an input amplitude decay rate $\kappa_{in} = \mathrm{FSR} / 4\mathcal{F} \simeq 2\pi\times\SI{8.3\pm0.1}{\kilo\hertz}$, where the error is dominated by the uncertainty in the cavity length.
The finesse of the cavity reduces to $\mathcal{F} = \num{12463\pm13}$ when the membrane--sandwich is placed in the cavity.
Such finesse corresponds to a cavity amplitude decay rate $\kappa  = \mathrm{FSR} / 2\mathcal{F} \simeq 2\pi\times\SI{66.8\pm0.8}{\kilo\hertz}$. The uncertainty in  $\omega_m$ is estimated from the fit of the mechanical mode [see Fig.~\ref{fig:Cumulative_sup}(a)], and in $\omega_b$ from the datasheet of the function generator (Agilent 33220A): it has a frequency resolution of \SI{1}{\micro\hertz} and an accuracy of \SI{0.1}{ppm\per day}, and is hence negligible for our analysis. From this we evaluate $\calK = \num{5.65\pm0.14}$. 

The mode--matching factor, that is the ratio between the reflected pump beam power in-- and out--of--resonance, has been estimated to be $\epsilon = \num{0.249\pm0.05}$, In fact, only a factor $\epsilon P_{in}$ is useful for coupling to the cavity mode $\mathrm{TEM_{00}}$. The powers we provide in the paper are rescaled by the mode--matching factor.
\begin{table}[h!]
	\caption{Optomechanical parameters used for the evaluation of $g_0$ by means of VSN integrals ($\Delta V^2$).}
	\centering
	\begin{tabular}{cccc}
		\hline		
		$\omega_{m}$  & & $2\pi\cdot\SI{229.753\pm 0.0001}{\kilo\hertz}$ \\
		$\omega_{b}$ & & $2\pi\cdot\SI{237.0}{\kilo\hertz}$ \\
		$\Delta$ & & $\SI{0}{\hertz}$ \\
		$\gamma_{m}$ & & $2\pi\cdot\SI{1.64\pm 0.1}{\hertz}$ \\
		$\calK$ & & $\num{5.65\pm0.14}$ \\
		$\beta$ & & $\SI{19.5\pm0.1}{\milli\radian}$ \\
		$\Delta V_m^2$ & & $\SI{2.8\pm0.4 e-10}{\square\volt}$ \\
		$\Delta V_b^2$ & & $\SI{1.669\pm0.001 e-8}{\square\volt}$ \\
		$\kappa$ & & $2\pi\cdot\SI{66.8\pm8}{\kilo\hertz}$ \\
		$\kappa_{in}$ & & $2\pi\cdot\SI{8.3\pm 0.1}{\kilo\hertz}$ \\
		$\epsilon$ & & $\num{0.249\pm0.05}$ \\
		${\rm FSR}$ & &  $2\pi\cdot\SI{1.67\pm0.02}{\giga\hertz}$ \\
		${\rm \calF}$ & &  $2\pi\cdot\num{12463\pm 13}$ \\
		$\lambda_0$ & & $\SI{1064}{\nano\meter}$ \\
		\hline
	\end{tabular}
	\label{tb:tab1}
\end{table}

\section{Calibration of $\beta$}
\label{App:A1}

\begin{figure}[b!]
	\centering
	\includegraphics[width=0.95\linewidth]{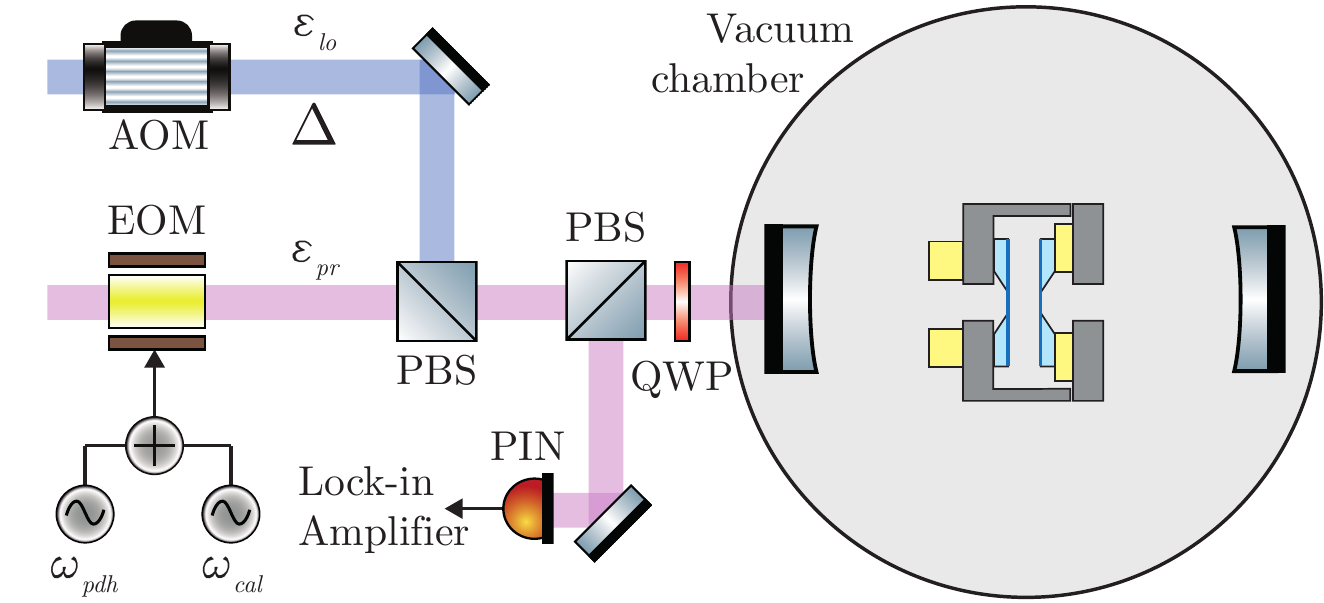}
	\caption{\new{Experimental setup for the characterization of the EOM. Two beams frequency detuned by $\Delta$ by means of an AOM impinge on a cavity out of resonance. The function generator HP~8657B used for  driving the AOM has a frequency resolution of \SI{1}{\hertz} and an accuracy of \SI{2}{ppm\per year}, hence negligible in our case.
	 The beams both reflected, interfere on the PBS before the cavity, and are finally collected on a PIN photodiode Thorlabs PDA10CF that has a flat bandwidth up to $\SI{150}{\mega\hertz}$. The signal is analysed with a lock--in amplifier. The power of the probe and pump beams are \SI{17}{\micro\watt}, and \SI{350}{\micro\watt}, respectively. The modulation $\omega_{pdh}$ and $\omega_{cal}$ are generated with two different function generators, a Siglent SDG 2122A and an Agilent 33220A, respectively. The two signals are then combined with a Minicircuit ZFSC-2-6+ and sent to the EOM with $\sim\SI{6}{\meter}$ BNC cable. QWP is a quarter--waveplate, and PBS a polarizing beam-splitter.}}
	\label{fig:cavitysetup_sup}
\end{figure}
The optical scheme used for the characterization of an electro--optic--modulator (EOM) is shown in Fig.~\ref{fig:cavitysetup_sup}. The probe beam $\varepsilon_{pr}$ goes through the EOM and is modulated at the frequency $\omega_{pdh}$ with modulation depth $\beta$:
\begin{align}
\label{eq:eps_probe_sup}
\varepsilon_{pr} &=  \varepsilon_0 \ee^{\ii\omega_L t + \ii\beta\sin(\omega_{pdh}t)}
\,.
\end{align}
Then $\varepsilon_{pr}$ is reflected by the optomechanical cavity, which is out of resonance and it is then collected on a PIN photodiode. Also the pump beam, $\varepsilon_{pm}$, detuned in frequency by $\Delta$ with respect to the probe by means of an acusto--optic--modulator (AOM), impinges on the cavity and is reflected:
\begin{align}
%\label{eq:eps_probe}
%\varepsilon_{pr} &=  \varepsilon_0 \ee^{\ii\omega_L t + \beta\sin(\omega_{pdh}t)} \\
\varepsilon_{lo} &=  \varepsilon_1 \ee^{\ii(\omega_L + \Delta)t}
\label{eq:eps_pump_sup}
\,.
\end{align}

\begin{figure}[b!]
	\centering
	\includegraphics[width=0.995\linewidth]{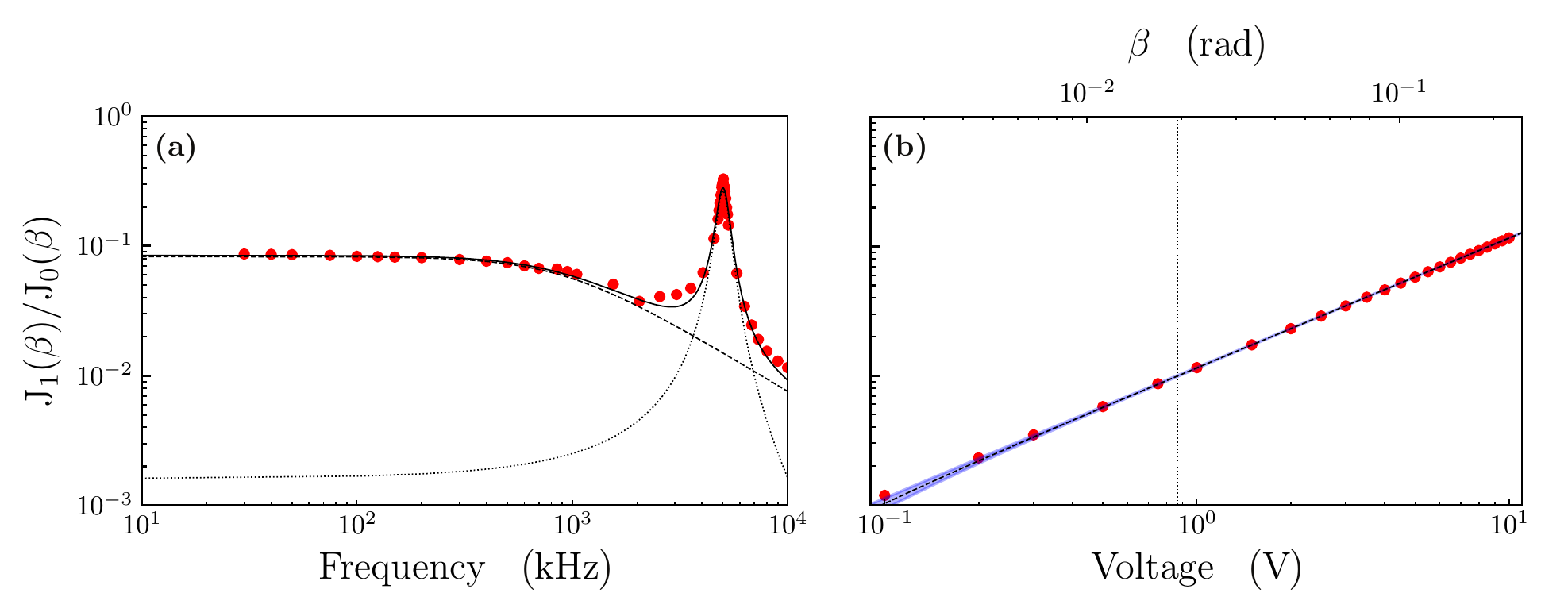}
	\caption{\textbf{(a)} Ratio $J_1(\beta)/J_0(\beta)$ as a function of the frequency at which the EOM is driven. The RLC resonant frequency is around \SI{5.015}{\mega\hertz}, and the Q--factor is \num{6.4}. The low pass filter at low frequencies with cut--off frequency around \SI{918.7}{\kilo\hertz} is probably due to some parasitic components in the RLC circuit. The black--dashed line represents the low--pass filter transfer function, the black--dotted line is the RLC transfer function and the black--solid line is the sum of the dahsed and dotted lines. \textbf{(b)} Ratio $J_1(\beta)/J_0(\beta)$ as a function of the voltage applied to the EOM at the a frequency of $\SI{237}{\kilo\hertz}$, which is the frequency of the beating note to calibrate the optomechanical spectra. The transduction coefficient is $\SI{22.5 \pm 0.1}{\milli\radian/\volt}$. The black--dotted line indicates our operating value which is \SI{19.5 \pm 0.1}{\radian}. The dashed--black line is a best--fit and the blue shadow represents the error in the estimation of the modulation depth.}
	\label{fig:pockelcellnew_sup}
\end{figure}

The polarization of the two beams is transformed such that they interfere on the last PBS. The interference signal is then collected by the PIN photodiode and analysed with a lock--in amplifier. 
We can expand Eq.~\eqref{eq:eps_probe_sup} in terms of Bessel's functions using the Jacobi--Anger expansion:
\begin{align}
\varepsilon_{pr} & = \varepsilon_0  \ee^{\ii\omega_L t}\sum_{n=-\infty}^{+\infty} J_n(\beta)\ee^{\ii n\omega_{pdh}t}
\label{eq:eps_probe_bessel_sup}
\,.
\end{align}
The interference of the two beams gives:
\begin{align}
%\nonumber
|\varepsilon_{pr} + \varepsilon_{lo}|^2 %= & \varepsilon_{0}^2 + \varepsilon_{1}^2  + 2\varepsilon_{0}\varepsilon_{1}\Real\left\{\sum_{n=-\infty}^{+\infty}J_n(\beta)\ee^{\ii n \omega_{pdh} t} \ee^{-\ii \Delta t}\right\} = \\
= & \mathrm{DC} + 2\varepsilon_{0}\varepsilon_{1} \sum_{n=-\infty}^{+\infty}J_n(\beta)\Re\left[\ee^{\ii (n \omega_{pdh} - \Delta) t}\right]
\label{eq:interference_sup}	
\,,
\end{align}
where $\mathrm{DC} = \varepsilon_{0}^2 + \varepsilon_{1}^2$. For $\beta \ll 1$, only the first order of the expansion of Eq.~\eqref{eq:interference_sup} is relevant:  
\begin{align}
	|\varepsilon_{pr} + \varepsilon_{lo}|^2 \sim  \mathrm{DC} 
	&+ 2\varepsilon_{0}\varepsilon_{1}\biggl\{ J_0(\beta)\Re\left[\ee^{-\ii\Delta t}\right] +
	\nonumber \\ 
	&+ J_1(\beta)\Re\left[\ee^{\ii(\omega_{pdh} - \Delta)t}\right] +
	 \\ 
	 &+ J_{-1}(\beta)\Re\left[\ee^{-\ii(\omega_{pdh} + \Delta)t}\right]\biggr\} + o(2\omega_{pdh})
	\nonumber
	\label{eq:interference_approx_sup} 
	\,.
\end{align}
The above equation shows that in the frequency domain the signal has a component at $\Delta$ proportional to $J_0(\beta)$, and two sidebands at frequencies $\omega_{pdh} - \Delta$ and $\omega_{pdh} + \Delta$ that are proportional to $J_1(\beta)$ and $J_{-1}(\beta)$, respectively. The ratio between the sideband term and the carrier term allows us to determine the modulation depth of the Pockels cell.

The Pockels cell we use in the experiment is a broadband New Focus 4004 model. To drive electronically the Pockels cell, we use an LC circuit. The Pockels cell acts like a capacitor with an impedance capacitance of $\sim\SI{20}{\pico\farad}$. To have the sideband frequencies far from the frequency of the carrier beam, we put in parallel to the capacitance a $\SI{50}{\micro\henry}$ inductance in order to get a resonant frequency around \SI{5}{\mega\hertz}. To characterize the LC circuit, we acquire the interference signal between $\varepsilon_{pr}$ and $\varepsilon_{lo}$ at the PIN photodiode, as shown in Fig. \ref{fig:cavitysetup_sup}. 
The signal is then demodulated by the lock--in amplifier at the frequency $\Delta$ for the carrier, and $\omega_{pdh} - \Delta$ and $\omega_{pdh} + \Delta$ for the lower and upper sidebands, respectively.  In Fig. \ref{fig:pockelcellnew_sup}(a) the ratio of the demodulated signals is plotted as a function of the modulation frequency $\omega_{pdh}$ applied to the EOM. The RLC circuit is centred around \SI{5.015}{\mega\hertz}, and the FWHM is about \SI{780}{\kilo\hertz}. From these values we get a Q--factor of  $\sim\num{6.4}$. The low pass filter at low frequencies with cut--off frequency around \SI{918.6}{\kilo\hertz} is probably due to some parasitic components in the RLC circuit. In Fig. \ref{fig:pockelcellnew_sup}(b), instead, is shown the ratio of the sidebands and the carrier areas obtained by integrating the lock--in amplifier signal over a bandwidth of $\SI{20}{\hertz}$, as a function of the voltage applied to the EOM at $\SI{237}{\kilo\hertz}$. The transduction coefficient from voltage to modulation depth at the RLC resonance is $\SI{22.5 \pm 0.1}{\milli\radian/\volt}$. Our operating value is \SI{19.5 \pm 0.1}{\radian}. This number is useful for the calibration of the mechanical spectra of the membranes inside the optomechanical cavity, when a calibration tone is sent via the EOM.

\section{Estimation of the area}
\label{App:A2}

\begin{figure}[hb!]
	\centering
	\includegraphics[width=0.975\linewidth]{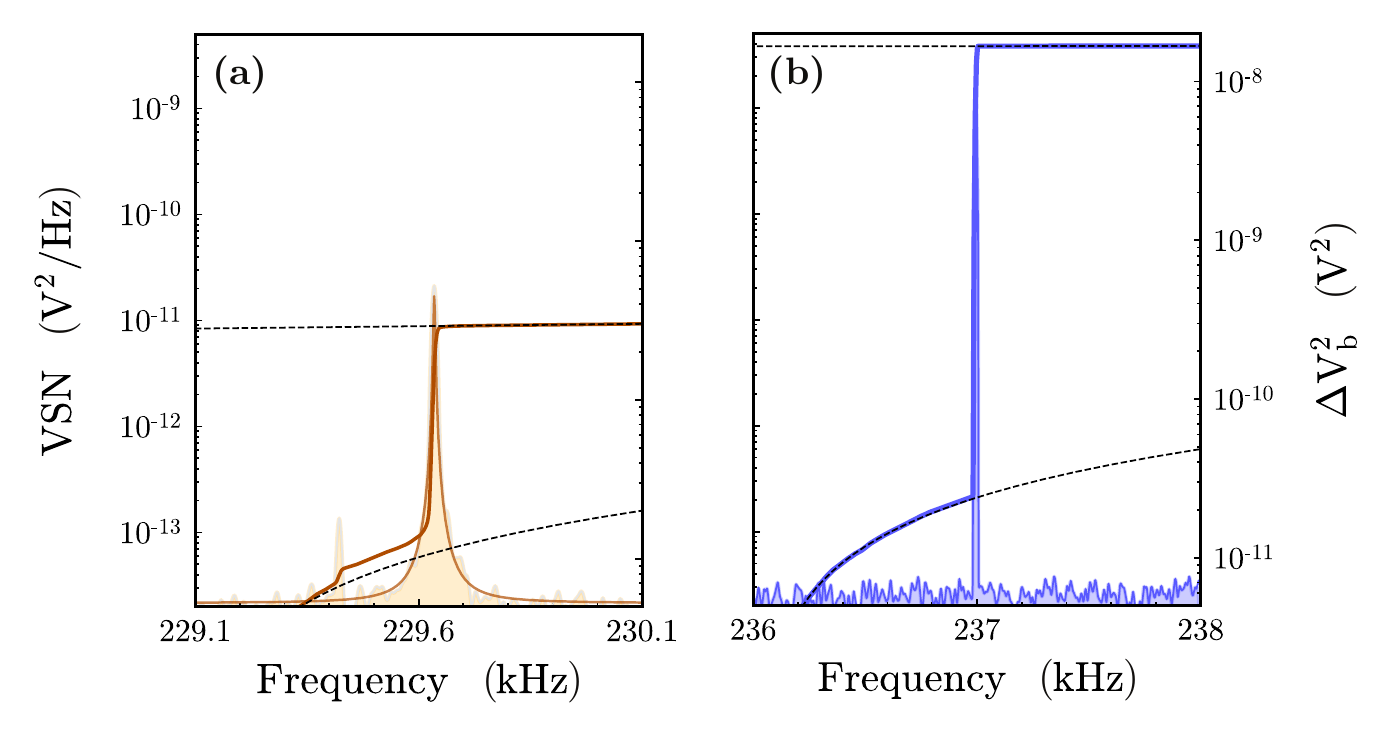}
	\caption{\textbf{(a)} VSN of the mechanical spectrum (light--yellow area) and cumulative function (brown line) for the mechanical mode. The light--brown line is a Lorentzian fit of the mechanical mode. The black--dashed lines are linear fit of the bottom and top of the cumulative function. The dotted line represents the distance between the two linear fits and gives an area of $\Delta V^2_m  = \SI{2.8\pm0.4 e-10}{\square\volt}$. \textbf{(b)} VSN of the calibration peak (light--blue area) and cumulative function (blue line) for the mechanical mode. The black--dashed lines are linear fit of the bottom and top of the cumulative function. The dotted line represents the distance between the two linear fit and gives an area of $\Delta V^2_b  = \SI{1.669\pm0.001 e-8}{\square\volt}$.}
	\label{fig:Cumulative_sup}
\end{figure}

The area of the mechanical spectra and the calibration tone is estimated by the cumulative function of the VSN for the mechanical mode, and the calibration tone (see Fig.~\ref{fig:Cumulative_sup}). The upper and lower parts of the cumulative function are fitted with a line (black--dashed line) and the distance between
the two linear fits is then estimated. The areas are $\Delta V^2_m  = \SI{2.8\pm0.4 e-10}{\square\volt}$, and $\Delta V^2_b  = \SI{1.669\pm0.001 e-8}{\square\volt}$, for the mechanical mode and the calibration tone, respectively. 

\section{Noninvasive detection}
\label{App:B}

We observe, that for $\Delta_j = 0$, the quantity $\Sigma$ of Eq.~\eqref{eq:Sigmaj_sup} is always equal to zero for any $\xi$. The key point here is that the radiation pressure term is exactly zero, at all orders, if the beam is kept at resonance. In fact, at resonance, for every term $n$ of the infinite sum in Eq.~\eqref{eq:Sigmaj_sup}, there is the term with $n + 1 = -l$, which is its opposite, due to the fact that $J_{-l}(x) = (-1)^lJ_l(x)$. 
%%
%\begin{eqnarray}
%	\Sigma(\xi,\kappa_i,0) 
%			 &=& \sum_{n =0}^{\infty}\frac{J_n\left(-\xi\right)J_{n+1}\left(-\xi\right)}
%						{[\ii n{\omega_{m}} + \kappa_j][-\ii (n+1)\omega_{m} + \kappa_j]} + 
%				\sum_{n =-1}^{-\infty}\frac{J_n\left(-\xi\right)J_{n+1}\left(-\xi\right)}
%						{[\ii n{\omega_{m}} + \kappa_j][-\ii (n+1)\omega_{m} + \kappa_j]}
%						\nonumber \\
%			 &=& \sum_{n =0}^{\infty}\frac{J_n\left(-\xi\right)J_{n+1}\left(-\xi\right)}
%						{[\ii n{\omega_{m}} + \kappa_j][-\ii (n+1)\omega_{m} + \kappa_j]} + 
%				\sum_{l=0}^{\infty}\frac{J_{-(l+1)}\left(-\xi\right)J_{-l}\left(-\xi\right)}
%						{[-\ii(l+1) {\omega_{m}} + \kappa_j][\ii l\omega_{m} + \kappa_j]}
%						\nonumber \\
%			 &=& \sum_{n =0}^{\infty}\frac{J_n\left(-\xi\right)J_{n+1}\left(-\xi\right)}
%						{[\ii n{\omega_{m}} + \kappa_j][-\ii (n+1)\omega_{m} + \kappa_j]} - 
%				\sum_{l=0}^{\infty}\frac{J_{l+1}\left(-\xi\right)J_{l}\left(-\xi\right)}
%						{[-\ii(l+1) {\omega_{m}} + \kappa_j][\ii l\omega_{m} + \kappa_j]} = 0
%	\,.\label{eq:Sigmaj_sup}
%\end{eqnarray}
%%
Therefore, we reach the important conclusion that the optical mode realises a perfectly noninvasive detection of the mechanical mode and of its nonlinearity, without any backaction, as it occurs in a Michelson interferometer readout, as long as it remains resonant with the cavity.

\section{Displacement calibration}
\label{App:C}

\begin{figure}[hb!]
	\begin{center}
		{\includegraphics[width=.315\textwidth]{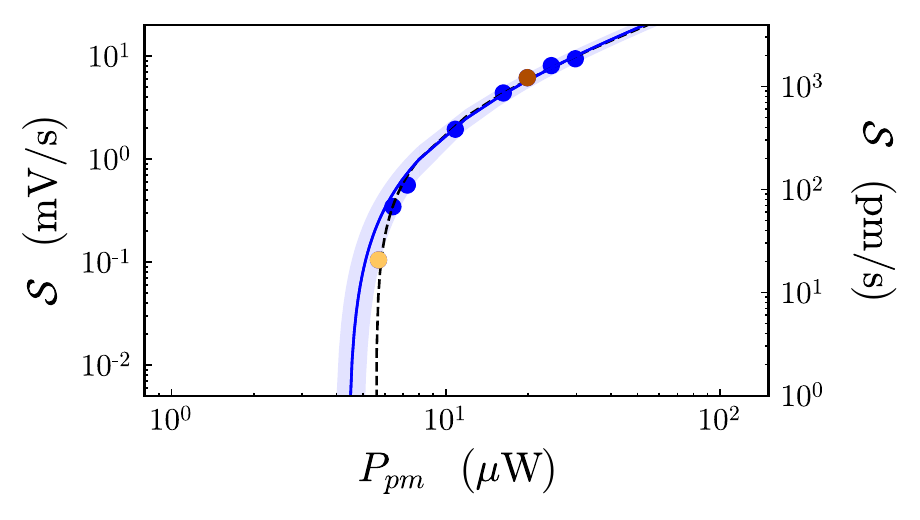}}
		\caption{
		Maximum slope in $\si{\volt\per\second}$  as a function of the pump power, $P_{pm}$, reported in Fig.~\ref{fig:Slope_LOG} of the main text.
		The right axis shows $\calS_x$ measured in $\si{\pico\meter\per\second}$, and it is calibrated by matching the voltage signal $V_m$ (first $\SI{5}{\second}$ in Fig.~\ref{fig:Figure_Time} of the main text) to the thermal displacement of the mechanical oscillator $q_T = \sqrt{2\bar n_m}\,x_{\rm zpf}$.
		}
		\label{figSM:Slope_LOG_sup}
	\end{center}
\end{figure}

The calibration in terms of displacement spectral noise (DSN) from the voltage spectral noise requires the knowledge of the effective mass through the mechanical zero-point fluctuation amplitude $x_\mathrm{zpf} = \sqrt{\hbar/2m_{eff}\omega_m}$. 
In fact, this is done by comparing the thermal displacement uncertainty of the mechanical oscillator at temperature $T$, given by $q = \sqrt{2\bar n_m}\,x_\mathrm{zpf}$, to the voltage displacement, either by integrating the VSN or through the time amplitude fluctuation. 
The determination of the effective mass of the mechanical modes, $m_{eff}$, is usually predetermined by numerical finite–element analysis. For the membrane we used in the experiment, we have $m_{eff} = \SI{1.74e-10}{\kilogram}$. The slope of the dynamics of the mechanical resonator when it undergoes a Hopf bifurcation and is driven into a limit cycle by the radiation-pressure-like interaction, calibrated from \si{\volt\per\second} to $\si{\pico\meter\per\second}$, is shown in Fig.~\ref{figSM:Slope_LOG_sup}(b), and it is obtained by matching the voltage signal $V_m$ (first $\SI{5}{\second}$ in Fig.~\ref{fig:Figure_Time} of the main text) to the thermal displacement of the mechanical oscillator.


\begin{thebibliography}{35}

\bibitem{Berman}
	P. R. Berman,
	Cavity quantum electrodynamics, 
	Boston, Academic Press, (1994).

\bibitem{rmp}
	M. Aspelmeyer, T. J. Kippenberg, and F. Marquardt, 
	Cavity optomechanics,
	{\it Rev. Mod. Phys.} 
	{\bf 86}, 1391 (2014).

\bibitem{Gorodetsky:2010uq}
	M. L. Gorodetsky, A. Schliesser, G. Anetsberger, S. Deleglise, and T. J. Kippenberg,
	Determination of the vacuum optomechanical coupling rate using frequency noise calibration, 
	{\it Optics Express\/} 
	{\bf 22}, 23236--23246 (2010).

\bibitem{Regal:2015}
	T. P. Purdy, P.-L. Yu, N. S. Kampel, R. W. Peterson, K. Cicak, R. W. Simmonds, and C. A. Regal,
	Optomechanical Raman-ratio thermometry,
	{\it Phys. Rev. A} 
	{\bf 92}, 031802(R) (2015).

\bibitem{Nielsen:2017}
	W. H. P. Nielsen, Y. Tsaturyan, C. B. Møller, E. S. Polzik, and A. Schliesser,
	Multimode optomechanical system in the quantum regime,
	{\it Proc. Natl. Acad. Sci. U.S.A.} 
	{\bf 114}, 62–66 (2017).
	
\bibitem{Rossi:2019} 
	M. Rossi, D. Mason, J. Chen, Y. Tsaturyan, and A. Schliesser,
	Measurement-based quantum control of mechanical motion, 
	{\it Nature}
	{\bf 563}, 53 (2018).

\bibitem{Karuza:2013}
	M. Karuza, C. Biancofiore, M. Bawaj, C. Molinelli, M. Galassi, R. Natali, P. Tombesi, G. Di~Giuseppe, and D. Vitali,
	Optomechanically induced transparency in a membrane-in-the-middle setup at room temperature, 
	{\it Phys. Rev. A} 
	{\bf 88}, 013804 (2013).

\bibitem{Regal:2013}
	T. P. Purdy, R. W. Peterson, and C. A. Regal,
	Observation of radiation pressure shot noise on a macroscopic object,
	{\it Science}
	{\bf 339}, 6121 801--804 (2013).

\bibitem{Painter:2019}
	G. S. MacCabe, H. Ren, J. Luo, J. D. Cohen, H. Zhou, A. Sipahigil, M. Mirhosseini, and O. Painter,
	Nano-acoustic resonator with ultralong phonon lifetime,
	{\it Science}
	{\bf 370}, 6518 840--843 (2020).

\bibitem{Painter:2020}
	H. Ren, M. H. Matheny, G. S. MacCabe, J. Luo, H. Pfeifer, M. Mirhosseini, and O. Painter,
	Two-dimensional optomechanical crystal cavity with high quantum cooperativity,
	{\it Nat. Commun.}
	{\bf 11}, 3373 (2020).
	
\bibitem{Marquardt2006}  
	F. Marquardt,  J. G. E. Harris, and S. M. Girvin, 
	Dynamical multistability induced by radiation pressure in high--finesse micromechanical optical cavities, 
	{\it Phys. Rev. Lett.} 
	{\bf 96}, 103901 (2006).

\bibitem{Carmon}
	T. Carmon, H. Rokhsari, L. Yang, T. J. Kippenberg, and K. J. Vahala,
	Temporal behavior of radiation--pressure--induced vibrations of an optical microcavity phonon mode,
	\textit{Phys. Rev. Lett.} 
	\textbf{94}, 223902 (2005).

\bibitem{Kippenberg2005}
	T. J. Kippenberg, H. Rokhsari, T. Carmon, A. Scherer, and K. J. Vahala,
	Analysis of radiation--pressure induced mechanical oscillation of an optical microcavity,
	\textit{Phys. Rev. Lett.} 
	\textbf{95}, 033901 (2005).

\bibitem{Metzger}
	C. Metzger, M. Ludwig, C. Neuenhahn, A. Ortlieb, I. Favero, K. Karrai, and F. Marquardt,
	Self--induced oscillations in an optomechanical system driven by bolometric backaction,
	\textit{Phys. Rev. Lett.} 
	\textbf{101}, 133903 (2008).

\bibitem{Krause:2015aa}
	A. G. Krause, J. T. Hill, M. Ludwig, A. H. Safavi-Naeini, J. Chan, F. Marquardt, and O. Painter,
	Nonlinear radiation pressure dynamics in an optomechanical crystal,
	\textit{Phys. Rev. Lett.} 
	{\bf 115}, 233601 (2015).

\bibitem{Buks2019}
	E. Buks, and I. Martin, 
	Self-excited oscillation and synchronization of an on-fiber optomechanical cavity,
	\textit{Phys. Rev. E} 
	\textbf{100}, 032202 (2019).

\bibitem{Siegman1996}
	A. E. Siegman, Lasers, University Science Books, Sausalito, CA, 
	(1986).

\bibitem{Josephson}
	S. H. Strogatz, Nonlinear Dynamics And Chaos: With Applications To Physics, Biology, Chemistry, And Engineering, 2nd edition, Westview Press, Boulder (2015).

\bibitem{Holmes2012} 
	C. A. Holmes,  C. P. Meaney, and G. J. Milburn,
	Synchronization of many nanomechanical resonators coupled via a common cavity field,  
	{\it Phys. Rev. E} 
	{\bf 85}, 066203 (2012).

\bibitem{Li:2020aa}
	W. Li, P. Piergentili, J. Li, S. Zippilli, R. Natali, and N. Malossi, and G. Di~Giuseppe, and D. Vitali,
	Noise robustness of synchronization of two nanomechanical resonators coupled to the same cavity field,
	{\it Phys. Rev. A} 
	{\bf 101}, 013802 (2020).

%\bibitem{Piergentili2020}
%	P. Piergentili, W. Li, R. Natali, N. Malossi, D. Vitali, and G. Di~Giuseppe,
%	Two--membrane cavity optomechanics: non-linear dynamics,
%	{\it arXiv:2009.04694}
%	(2020).

\bibitem{Piergentili2020}
	P. Piergentili, W. Li, R. Natali, N. Malossi, D. Vitali, and G. Di~Giuseppe,
	Two-membrane cavity optomechanics: non-linear dynamics,
	{\it 	arXiv:2009.04694}
	(2020).
	
\bibitem{Piergentili:2018aa}
	P. Piergentili, L. Catalini, M. Bawaj, S. Zippilli, N. Malossi, R. Natali, D. Vitali, and G. Di~Giuseppe,
	Two--membrane cavity optomechanics, 
	{\it New J. Phys.} 
	{\bf 20}, 083024 (2018).

\bibitem{Rossi:2017aa}
	M. Rossi, N. Kralj, S. Zippilli, R. Natali, A. Borrielli, G. Pandraud, E. Serra, G. Di Giuseppe, and D. Vitali,
	Enhancing sideband cooling by feedback--controlled light,
	{\it Phys. Rev. Lett.}
	{\bf 119}, 123603 (2017).

\end{thebibliography}
\end{document}